# Topological Defects Coupling Smectic Modulations to Intra-unit-cell Nematicity in Cuprate

A. Mesaros[*,1,2], K. Fujita[*,2,3,4], H. Eisaki[5], S. Uchida[4], J.C. Davis[2,3,6], S. Sachdev[7], J. Zaanen[1], M.J. Lawler[2,8] and Eun-Ah Kim[2]

1. Instituut-Lorentz for Theoretical Physics, Universiteit Leiden, The Netherlands.
2. Laboratory for Atomic and Solid State Physics, Department of Physics, Cornell University, Ithaca, NY 14853, USA.
3. Condensed Matter Physics and Materials Science Department, Brookhaven National Laboratory, Upton, NY 11973, USA.
4. Department of Physics, University of Tokyo, Bunkyo-ku, Tokyo 113-0033, Japan.
5. Institute of Advanced Industrial Science and Technology, Tsukuba, Ibaraki 305-8568, Japan.
6. School of Physics and Astronomy, University of St. Andrews, North Haugh, St. Andrews, Fife KY16 9SS, Scotland.
7. Department of Physics, Harvard University, Boston, MA 02138, USA.
8. Department of Physics, Applied Physics and Astronomy, Binghamton University, Binghamton, NY 13902-6000, USA.

*These authors contributed equally to this work

We study the coexisting smectic modulations and intra–unit-cell nematicity in the pseudogap states of underdoped $Bi_2Sr_2CaCu_2O_{8+\delta}$. By visualizing their spatial components separately, we identified $2\pi$ topological defects throughout the phase-fluctuating smectic states. Imaging the locations of large numbers of these topological defects simultaneously with the fluctuations in the intra–unit-cell nematicity revealed strong empirical evidence for a coupling between them. From these observations, we propose a Ginzburg-Landau functional describing this coupling and demonstrate how it can explain the coexistence of the smectic and intra–unit-cell broken symmetries and also correctly predict their interplay at the atomic scale. This theoretical perspective can lead to unraveling the complexities of the phase diagram of cuprate high-critical-temperature superconductors.

**1**     Electronic liquid crystals are proposed to occur when the electronic structure of a material breaks the spatial symmetries of its crystal lattice (1–8). In theory, nematic electronic liquid crystals would preserve the lattice translational symmetry but break the discrete rotational symmetry, whereas smectic (striped) electronic liquid crystals would break both. These concepts have played an important role in theoretical considerations of the pseudogap phase of underdoped cuprates (1–8).

**2** At hole densities ($p$) below ~16%, cuprates exhibit d-wave superconductivity at lowest temperatures and the pseudogap phase above the superconductor's critical temperature, $T_c$. Although it is unknown which broken symmetries (if any) cause the pseudogap phase, both nematic and smectic broken symmetry states have been reported in different underdoped cuprate compounds (9–18). Spin and charge smectic broken symmetry (stripes) exists in $La_{2-x-y}Nd_ySr_xCuO_4$ and in $La_{2-x}Ba_xCuO_4$ when x ~ 0.125 (6, 9–12). On the other hand, broken 90°-rotational symmetry is reported in underdoped $YBa_2Cu_3O_{6+\delta}$ (13, 15–17), underdoped $Bi_2Sr_2CaCu_2O_{8+\delta}$ (14, 18), and underdoped $HgBa_2CuO_{4+x}$ (19). These states are highly distinct: The former breaks both translational symmetry with a finite wave vector $\vec{q} = \vec{S}$, where the magnitude of $\vec{S}$ is the wave number for the modulation, and 90°-rotational symmetry (9–12), whereas the latter is associated with intra–unit-cell breaking of 90°-rotational symmetry (15, 18–20). A key challenge is therefore to understand the interactions between these phenomena (9–27).

**3** We consider the coexisting smectic modulations and intra–unit-cell nematicity in the pseudogap energy electronic structure of the underdoped high-$T_c$ superconductor $Bi_2Sr_2CaCu_2O_{8+\delta}$ (18, 20) by using approaches derived from studies of classical liquid crystals. In those systems, fluctuating nematic and disordered smectic states coexist, and their dominant coupling can be captured successfully by using Ginzburg-Landau theory (22, 24, 25). The influence of $2\pi$ phase-winding topological defects of the smectic was key to those studies. But the extension of such classical ideas to electronic systems presents some new challenges. First, the intra–unit-cell $C_4$-breaking observed at nanoscale in the cuprate pseudogap states (18, 20) is distinct from nematicity in a classical liquid crystal, because it has Ising symmetry resulting from the existence of the crystal lattice. Moreover, whether $2\pi$ topological defects even exist within the cuprate pseudogap smectic states was unknown.

*4* Topological defects are the fundamental emergent excitations when a new ordered phase is formed by breaking a continuous symmetry (21, 22). They are singular points or lines in the otherwise spatially continuous configuration of the order-parameter field. For example, when the order-parameter field is a complex function $\Psi(\vec{r}) = \Psi_0 e^{i\varphi(\vec{r})}$ of the position $\vec{r}$, the phase $\varphi(\vec{r})$ winds by integer multiples of $\pm 2\pi$ around every topological defect. Classic examples include the quantized vortices in bosonic and fermionic superfluids (23) and the quantized fluxoids of superconductors (22, 23) (Fig. 1A inset). Systems with broken translational symmetry, such as crystals or smectic liquid crystals, also exhibit $2\pi$ phase-winding topological defects. In a crystal, when a single line of atoms (Fig. 1A, black dots) terminates at an edge dislocation, nearby atoms are distorted away from their ideal lattice locations, resulting in a spatially varying phase of periodic modulations that winds around the dislocation core by precisely $2\pi$ (22). In smectic liquid crystals, the equivalent topological defects are referred to as (smectic) dislocations. Again, each dislocation core is surrounded by a region where the phase of the periodic (smectic) modulations (white lines in Fig. 1A) winds by exactly $2\pi$. These topological defects are uniquely important in classical liquid crystals because their properties reveal the dominant coupling between the nematic field and the smectic field. In fact, quasi–long-range smectic-*A* order in two dimensions is destroyed by this coupling, which lowers the energy cost of smectic dislocations, allowing their spontaneous appearance at any temperature (22, 24, 25). We apply an analogous theoretical approach to coexisting broken electronic symmetries in underdoped cuprates.

**5** Spectroscopic imaging scanning tunneling microscopy (SI-STM) allows visualization of electronic broken symmetries in cuprates (18, 20, 26, 27) by using atomically resolved spatial images of $Z(\vec{r}, E) = dI/dV(\vec{r}, E \equiv +eV)/dI/dV(\vec{r}, E \equiv -eV)$, where $dI/dV(\vec{r}, V)$ is the spatially resolved differential tunneling conductance [supporting online material (SOM) a]. In underdoped cuprates, energy-independent symmetry breaking is vivid in the nondispersive $Z(\vec{r}, E)$ modulations at the pseudogap energy scale $E \sim \Delta_1$ (18, 20, 26–28). The coexistence of intra–unit-cell nematicity and smectic modulations (18, 20) appears to be a robust property of these electronic structure images of the cuprate pseudogap states, being virtually identical in $Bi_2Sr_2CaCu_2O_{8+\delta}$ and $Ca_{2-x}Na_xCuO_2Cl_2$ (20) and unchanged from below to above $T_c$ (27).

**6** To separate the components of the $E \sim \Delta_1$ electronic structure, each $Z(\vec{r}, E)$ image is first distortion-corrected to render the atomic sites in a perfectly periodic array (18). Then, to deal with the spatial disorder in $\Delta_1(\vec{r})$, $E$ is rescaled locally to $e(\vec{r}) = E/\Delta_1(\vec{r})$, yielding $Z(\vec{r}, e)$; all the broken symmetry phenomena of the pseudogap states then occur together in a single image $Z(\vec{r}, e = 1)$ (Fig. 1B and SOM a). Then, when $\tilde{Z}(\vec{q}, e = 1)$, the Fourier transform of $Z(\vec{r}, e = 1)$, is calculated (Fig. 1B inset), it exhibits four salient features: the Bragg peaks at $\vec{q} = \vec{Q}_x$ and $\vec{Q}_y$ (red circles) and the smectic modulation peaks $\vec{q} = \vec{S}_x$ and $\vec{S}_y$ (blue circles). The phase-resolved Bragg-peak Fourier components can then be used to detect intra–unit-cell symmetry breaking within each $Z(\vec{r}, e = 1)$ image (18).

**7**   We focus on intra–unit-cell "nematicity" defined by $\langle O_n(e) \rangle \equiv \left( \mathrm{Re}\,\tilde{Z}(\vec{Q}_y, e) - \mathrm{Re}\,\tilde{Z}(\vec{Q}_x, e) \right) / \overline{Z}(e)$, where $\overline{Z}(e)$ is the spatial average of $Z(\vec{r}, e)$, as a measure of the observed inequivalence between *x*- and *y*-axis electronic structure within the CuO$_2$ unit cell (18, 20, 27). A finite $\langle O_n(e) \rangle$ implies that the C$_{4v}$ symmetry of an ideal CuO$_2$ plane has been reduced at most to C$_{2v}$ symmetry. There are eight symmetry reduction possibilities for a system with full C$_{4v}$ symmetry; finite $\langle O_n(e) \rangle$ further restricts this to four. Information regarding further symmetry lowering (such as inversion symmetry breaking) can determine the actual symmetry of pseudogap states, but those issues are beyond the scope of this paper. A coarse-grained image $O_n(\vec{r}, e = 1)$ representing the local inequivalence of *x*- and *y*-axis electronic structure (18) is presented in Fig. 1C. The panel shows how, although $O_n(\vec{r}, e = 1)$ is strongly fluctuating at the nanoscale in very underdoped Bi$_2$Sr$_2$CaCu$_2$O$_{8+\delta}$, it has a finite average value within such a field of view.

**8**   The quite distinct properties of the smectic electronic structure modulations at $E \sim \Delta_1$ can be examined independently of the intra–unit-cell symmetry breaking by focusing only on the incommensurate modulation peaks $\vec{S}_x$ and $\vec{S}_y$. A coarse-grained image of the local degree of smectic symmetry breaking $O_s(\vec{r}, e = 1)$ (Fig. 1D and SOM b) reveals the very short correlation length of the strongly disordered smectic (18, 20, 26–28). The amplitude and phase of two unidirectional modulation components (along *x*, *y*) within the box in Fig. 1B can be further extracted, as shown in Fig. 2, A and B (29). To do so, we denote the local contribution to the $\vec{S}_x$ modulations at position $\vec{r}$ by a complex field $\psi_1(\vec{r})$. This contributes to the $Z(\vec{r}, e = 1)$ data as

$$\psi_1(\vec{r})e^{i\vec{S}_x\cdot\vec{r}} + \psi_1^*(\vec{r})e^{-i\vec{S}_x\cdot\vec{r}} \equiv 2|\psi_1(\vec{r})|Cos(\vec{S}_x\cdot\vec{r} + \varphi_1(\vec{r})) \qquad (1)$$

thus allowing the local phase $\varphi_1(\vec{r})$ of $\vec{S}_x$ modulations to be mapped similarly for the local phase $\varphi_2(\vec{r})$ of the $\vec{S}_y$ modulations. In Fig. 2, C and D, we show images of $\varphi_1(\vec{r})$ and $\varphi_2(\vec{r})$ derived from $Z(\vec{r}, e=1)$. They reveal that the smectic phases $\varphi_1(\vec{r})$ and $\varphi_2(\vec{r})$ take on all values between 0 and ±2π in a highly complex spatial pattern. Even more important is the detection of a large number of topological defects with ±2π phase winding. These are indicated by black (+2π) and white (−2π) circles in Fig. 2, C and D, and occur in about equal numbers (as one might anticipate from the likely macroscopic energy cost of an uncompensated dislocation). A typical example of an individual topological defect (solid box in Fig. 2, A and C) is shown in Fig. 3, A and B. The dislocation core (Fig. 3B) and its associated 2π phase winding are clearly seen (Fig. 3A). Moreover the amplitude of $\psi_1(\vec{r})$ or $\psi_2(\vec{r})$ always approaches zero near each topological defect, as expected. These data are all in close agreement with the theoretical expectations for quantum smectic dislocations (Fig. 1A).

**9** Imaging the locations of these topological defects (Fig. 2, C and D) simultaneously with the intra–unit-cell nematicity (Fig. 1C) reveals another key result. Figure 4A shows the locations of all topological defects in Fig. 2, C and D, plotted as black dots on the simultaneously acquired image $\delta O_n(\vec{r}) \equiv O_n(\vec{r}) - \langle O_n \rangle$ representing the fluctuations of the intra–unit-cell nematicity. By eye, nearly all the topological defects appear located in (white) regions of vanishing $\delta O_n(\vec{r}) = 0$. This can be quantified by plotting the distribution of distances of topological defects from the nearest zero of $\delta O_n(\vec{r})$, thereby showing that they are far smaller than expected if the topological defects were uncorrelated with $\delta O_n(\vec{r})$ (Fig. 4A inset and SOM c).

These data provide empirical evidence for a coupling between the smectic topological defects and the fluctuations of the intra–unit-cell nematicity at $E \sim \Delta_1$.

**10**   To establish a Ginzburg-Landau (GL) model representing such a coupling, one needs to determine first whether the $\delta O_n(\vec{r})$ fluctuations are coupled to the phase or the amplitude of the smectic modulations (30–33). Whether the modulations are commensurate (periodic with wavelength rational multiple of $a_0$) or incommensurate is key. For incommensurate modulations, a smooth deformation of the phase (Fig. 3A) costs a vanishingly small energy, whereas phase fluctuations always cost a finite energy for commensurate modulations. On the other hand, fluctuations of the modulation amplitude (Fig. 3D) cost a finite energy in both cases (34). There are multiple reasons to conclude that we are dealing with incommensurate modulations. First, the locations of $\vec{S}_x$ and $\vec{S}_y$ are not necessarily at a commensurate point in $\vec{q}$ space (Fig. 1B, inset), and they change continuously with hole density (26) (Fig. 3C). More profoundly, a complex histogram of $\psi_1(\vec{r})$ or $\psi_2(\vec{r})$ (Fig. 3D) shows little predominant phase preference overall. At a few high-amplitude locations (Fig. 3D), there is a particular phase preference consistent with short range commensurate "nanostripes" (20). However the continuous winding around each defect (Fig. 2) is in clear contrast to discrete jumps when only specific values of phase are allowed (35). Hence, these observations support the incommensurate picture in which the smectic broken symmetry exhibits free winding of the phase. Thus, our third advance is the demonstration that the simultaneously broken electronic symmetries in the $E \sim \Delta_1$ states consist of intra–unit-cell nematicity coexisting with disordered and phase fluctuating smectic modulations.

**11**    Spatial patterns of coexisting smectic modulations and intra–unit-cell nematicity, as well as their coupling, may be described most naturally by a corresponding GL functional. For the locally fluctuating $\vec{S}_x$ modulations represented by $\psi_1(\vec{r})$, the GL functional is

$$F_{GL}[\psi_1(\vec{r})] = \int d^2r \left[ a_x |\nabla_x \psi_1(\vec{r})|^2 + a_y |\nabla_y \psi_1(\vec{r})|^2 + m|\psi_1(\vec{r})|^2 \right] \quad (2)$$

Here, $a_x \neq a_y$ and $m$ are phenomenological GL parameters [assuming $x$ and $y$ directions are inequivalent (18)]. $F_{GL}$ is a generalization of the GL free energy of a density modulation in one spatial dimension (22). It is similar to the GL free energy of a superfluid. As it is for superfluids, fluctuations in phase $\varphi_1(\vec{r})$ enter $F_{GL}$ only through the spatial derivative terms because

$$|\nabla_x \psi_1(\vec{r})|^2 = (\nabla_x |\psi_1(\vec{r})|)^2 + |\psi_1(\vec{r})|^2 (\nabla_x \varphi_1(\vec{r}))^2 \quad (3)$$

The absence of long-range smectic order (Figs. 1D and 2) despite the finite modulation amplitudes (except within dislocation cores) implies phase fluctuations play the predominant role in smectic disordering. Further, the finite density of topological defects (Fig. 2) also indicates that Eq. 2 cannot provide a complete description of the phenomena. This is because an isolated topological defect will cost an energy that grows as a logarithm of the system size and hence is unlikely to occur. Yet we observe large numbers of isolated ±2π topological defects (Fig. 2). Therefore, coupling to other degrees of freedom must reduce the energy of the smectic dislocations. For the case of a classical nematic liquid crystal on the verge of freezing into a smectic-*A*, de Gennes discovered (24) a GL free energy describing how the nematic fluctuations lower the energy cost of smectic dislocations to a finite amount, thus allowing for the isolated topological defects to appear and resulting in destruction of quasi– long-range smectic order in two dimensions (24). With such a historical guide, we consider the interplay between the intra–unit-cell nematicity and incommensurate smectic modulations by including $\delta O_n(\vec{r})$ fluctuations in the above GL functional.

**12**   When $\langle O_n \rangle \neq 0$ (Fig. 1C) (18), the local fluctuation $\delta O_n(\vec{r}) \equiv O_n(\vec{r}) - \langle O_n \rangle$ (Fig. 4A) is the natural small quantity to enter the GL functional [when $\langle O_n \rangle = 0$ possibly at higher dopings, the expansion should be in terms of $O_n(\vec{r})$ with the appropriate symmetry]. Coupling to the smectic fields can then occur either through phase or amplitude fluctuations of the smectic. Here, we focus on the former, which means that $\delta O_n(\vec{r})$ couples to local shifts of the wave vectors $\vec{S}_x$ and $\vec{S}_y$. Replacing the gradient in the *x* direction by a covariant-derivative-like coupling gives

$$\nabla_x \psi_1(\vec{r}) \to (\nabla_x + i c_x \delta O_n(\vec{r})) \psi_1(\vec{r}), \qquad (4)$$

and similarly for the gradient in the *y* direction, to yield a GL term coupling the nematic to smectic states. The vector $\vec{c} = (c_x, c_y)$ represents by how much the wave vector, $\vec{S}_x$, is shifted for a given fluctuation $\delta O_n(\vec{r})$. Hence, we propose a GL functional (for modulations along $\vec{S}_x$) based on symmetry principles and $\delta O_n(\vec{r})$ and $\psi_1(\vec{r})$ being small:

$$F_{\mathrm{GL}}[\delta O_n, \psi_1] = F_n[\delta O_n] + \int d^2 r \Big( a_x \big|(\nabla_x + i c_x \delta O_n)\psi_1\big|^2 + a_y \big|(\nabla_y + i c_y \delta O_n)\psi_1\big|^2 + m|\psi_1|^2 + \ldots \Big), \qquad (5)$$

where … refers to terms we can neglect for the present purpose (SOM d). If we were to replace $\vec{c}\,\delta O_n(\vec{r})$ by $\tfrac{2e}{\hbar}\vec{A}(\vec{r})$ where $\vec{A}(\vec{r})$ is the electromagnetic vector potential, Eq. 5 becomes the GL free energy of a superconductor; its minimization in the long-distance limit yields $\vec{A}(\vec{r}) = \frac{\hbar}{2e}\vec{\nabla}\varphi(\vec{r})$ and thus quantization of its associated magnetic flux (22, 23). Analogously, minimization of Eq. 5 implies $\delta O_n(\vec{r}) = \vec{l}\cdot\vec{\nabla}\varphi(\vec{r})$ surrounding each topological defect (SOM e). Here, the vector $\vec{l}$ is proportional to ($a_x$; $a_y$) and lies along the line where $\delta O_n(\vec{r}) = 0$. The resulting key prediction is that $\delta O_n(\vec{r})$ will vanish along the line in the direction of $\vec{l}$ that passes through the core of the topological defect, with $O_n(\vec{r})$ becoming greater on one side and less on the other (Fig. 4B). Additional coupling to the smectic amplitude can shift the location of the topological defect away from the line of $\delta O_n(\vec{r}) = 0$ (SOMe).

**13** To test whether this GL model correctly captures the observed (Fig. 4, A and B) $\delta O_n - \psi_s$ coupling in Bi$_2$Sr$_2$CaCu$_2$O$_{8+\delta}$, we extend Eq. 5 to include both $\vec{S}_x$ and $\vec{S}_y$ smectic modulations. We then simulate the profile of $\delta O_n(\vec{r})$, treating the phase and amplitude of smectic fields $\psi_1(\vec{r})$ and $\psi_2(\vec{r})$ (Fig. 2) as mean-field input that will determine $\delta O_n(\vec{r})$ according to Eq. 5 (SOM e). Figure 4, C and D, shows the overlay of topological defect locations within the small boxes in Fig. 4A ond $O_n(\vec{r})$ as simulated by using Eq. 5 (SOMe). This demonstrates directly how the GL functional associates fluctuations in $\delta O_n(\vec{r})$ with the smectic topological defect locations in the fashion of Fig. 4B. The close similarity between the measured $\delta O_n(\vec{r})$ in Fig. 4, E and F, and the simulation in Fig. 4, C and D, with cross-correlation coefficients of 56% and 62% demonstrates how the minimal GL functional of Eq. 5 captures the interplay between the measured $\delta O_n(\vec{r})$ fluctuations (Fig. 4A) and disordered smectic modulations (Fig. 2). And, as expected with extrinsic disorder (36), the GL parameters vary somewhat from location to location (SOM f ). Indeed, a simultaneous "gapmap" (SOM g) shows vividly how much additional (probably dopant-atom-related) disorder coexists with the phenomena analyzed here.

14   Our results can lead to advances in understanding of coexisting and competing electronic phenomena in underdoped cuprates (9–20). By identifying $2\pi$ topological defects within the phase-fluctuating smectic states and that they are associated with the spatial fluctuations of the robust intra–unit-cell nematicity (18, 20), we demonstrated empirically a coupling between these two locally broken electronic symmetries of the cuprate pseudogap states. This allowed identification of a GL functional that explains how these phenomena coexist and predicts their interplay at the atomic scale. For example, the GL model explains why it is possible for the intra–unit-cell nematicity to have finite average $\langle O_n(\vec{r})\rangle \neq 0$ (Fig. 1C) even though the smectic modulations are disordered (Figs. 2 and 3) (18). This is because $2\pi$ topological defects induce fluctuations of $\delta O_n(\vec{r})$ with respect to $\langle O_n(\vec{r})\rangle$, but the dislocation cores sit close to locations where $O_n(\vec{r}) = \langle O_n(\vec{r})\rangle$ and thus do not disrupt this state directly (SOM e). Perhaps most importantly, if the tendency for intra–unit-cell nematicity to coexist with a disordered electronic smectic demonstrated here is ubiquitous to underdoped cuprates, which broken symmetry manifests at the macroscopic scale (9–20) depends on the coefficients in the GL functional and on other material-specific aspects, such as crystal symmetry. Therefore, the GL model introduced here provides a good starting point to address these issues and, eventually, the interplay between the different broken electronic symmetries and the superconductivity.

**Figure Captions**

**Figure 1**

a. Schematic image of an edge dislocation in a crystalline solid (solid circles indicate atomic locations) and in the two-dimensional smectic phase of a liquid crystal (solid white lines indicate modulation period). In both cases, it is the spatial phase of periodic modulations that winds around the dislocation core by precisely $2\pi$. (Inset) Schematic image of a superfluid or superconducting vortex overlapped with its phase field, which winds by exactly $2\pi$.

b. Sub–unit-cell resolution image of the electronic structure at the pseudogap energy $Z(\vec{r}, e=1)$. (Inset) Its Fourier transform of $\tilde{Z}(\vec{q}, e=1)$, which demonstrates that the $\vec{q}$-space electronic structure contains two components, nematic [red circles at the Bragg peaks, see (18)] and smectic (blue circles). The smectic peaks are centered at $|\vec{S}_x| = |\vec{S}_y|$ = 0.72($2\pi/a_0$). White box is field of view (FOV) of Fig. 2, A and B. $T_c$ of the sample is 50 K.

c. Spatial variation of the electronic nematicity $O_n(\vec{r}, e=1)$ in the same FOV as in (B). (Inset) The Bragg peak intensities are compared along *x* and *y* directions.

d. Spatial variation of the smectic electronic structure modulations $O_s(\vec{r}, e=1)$ [see (18)].

**Figure 2**

a. Smectic modulations along *x* direction are visualized by Fourier filtering out all the modulations of $Z(\vec{r}, e=1)$ except those surrounding $\vec{S}_x$, in the FOV indicated by the broken boxes in Fig. 1B and in (C).

b. Smectic modulations along *y* direction are visualized by Fourier filtering out all the modulations of $Z(\vec{r},e=1)$ except those surrounding $\vec{S}_y$, in the FOV indicated by the broken boxes in Fig. 1B and in (D).

c(d). Phase field $\varphi_1(\vec{r})$ and $\varphi_2(\vec{r})$ for smectic modulations along *x* and *y* direction, respectively, exhibiting the topological defects at the points around which the phase winds from 0 to $2\pi$ (in the FOV same as in Fig. 1B). Depending on the sign of phase winding, the topological defects are marked by either white or black dots. The broken red circle is the measure of the spatial resolution determined by the cut-off length ($3\sigma$) in extracting the smectic field from $\tilde{Z}(\vec{q},e=1)$. We did not mark defect-antidefect pairs when they are tightly bound by separation distances shorter than the cut-off length scale.

**Figure 3**

a. Phase field around the single topological defect in the FOV indicated by the solid box in Fig. 2, A and C.
b. Smectic modulation around the single topological defect in the same FOV, showing that the dislocation core is indeed at the center of the topological defect and that the modulation amplitude tends to zero there. This is true for all the $2\pi$ topological defects identified in Fig. 2.
c. Doping dependence of the wavelength for the smectic modulations along wave vectors $\vec{S}_x$ and $\vec{S}_y$ (26).
d. Two-dimensional histogram of real and imaginary components of the measured smectic field $\psi_1(\vec{r})$.

**Figure 4**

a. Fluctuations of electronic nematicity $\delta O_n(\vec{r}, e=1)$ obtained by subtracting the spatial average $\langle O_n(\vec{r}, e=1) \rangle$ from $O_n(\vec{r}, e=1)$ (Fig. 1C). The simultaneously measured locations of all $2\pi$ topological defects are indicated as black dots. They are primarily found near the lines where $\delta O_n(\vec{r}, e=1) = 0$. (Inset) The distribution of distances between each topological defect and its nearest $\delta O_n(\vec{r}, e=1) = 0$ contour. This is compared to the expected average distance if there is no correlation between $\delta O_n(\vec{r}, e=1)$ and the topological defect locations. There is a very strong tendency for the distance to the nearest $\delta O_n(\vec{r}, e=1) = 0$ contour to be small. The boxes show regions that are blown up in (E) and (F) and compared to simulations in (C) and (D).

b. Theoretical $\delta O_n(\vec{r}, e=1)$ predicted by the GL model in Eq. 5 (top) at the site of a single smectic topological defect (bottom). The vector $\vec{l}$ lies along the zero-fluctuation line of $O_n(\vec{r}, e=1)$.

c,d $\delta O_n(\vec{r}, e=1)$ obtained by numerical simulation using Eq. 5 and the experimentally obtained topological defect configurations (black dots). Red broken circle is the measure of the spatial resolution determined by the cut-off length ($3\sigma$) in extracting the smectic field. (See SOM f for the details of the numerical simulation).

e,f Measured $\delta O_n(\vec{r}, e=1)$ in the fields of view of (C) and (D). The achieved cross correlation is well above the lower bound for statistical significance (SOM f).

**Acknowledgements**

We thank E. Fradkin, S. Kivelson M. Norman, J.P. Sethna and J. Tranquada, for helpful discussions and communications. Theoretical studies at Cornell were supported by NSF Grant DMR-0520404 to the Cornell Center for Materials Research and by NSF Grant DMR-0955822. Experimental studies are supported by the Center for Emergent Superconductivity, an Energy Frontier Research Center, headquartered at Brookhaven National Laboratory and funded by the U.S. Department of Energy, under DE-2009-BNL-PM015 as well as by a Grant-in-Aid for Scientific Research from the Ministry of Science and Education (Japan) and the Global Centers of Excellence Program for Japan Society for the Promotion of Science. The work at Harvard University was supported by DMR-0757145. The work at Leiden University was supported by the Nederlandse Organisatie voor Wetenschappelijk Onderzoek (NWO) through a Spinoza Award. AM is grateful for the hospitality of E.-A.K.

Figure 1

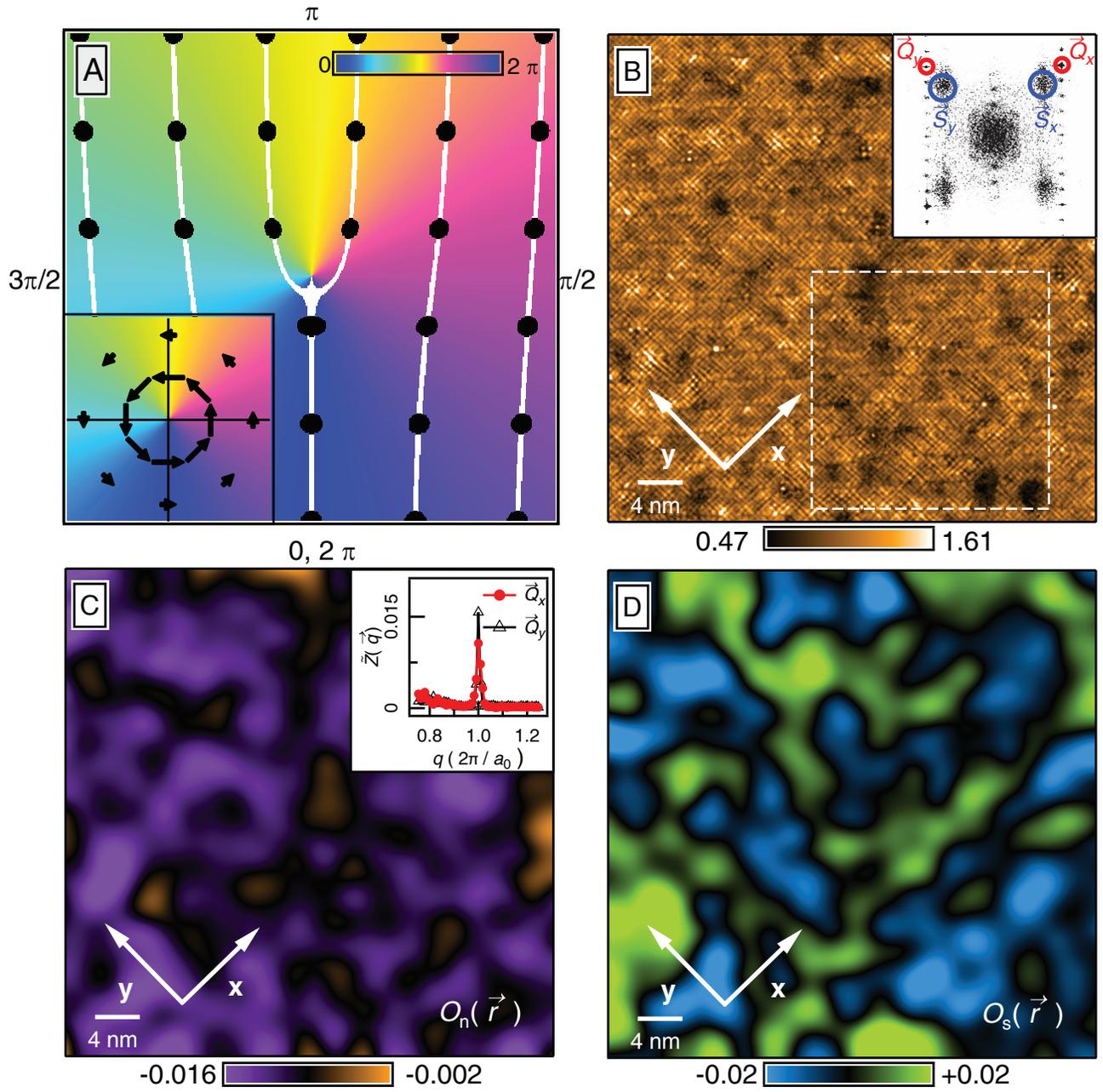

Figure 2

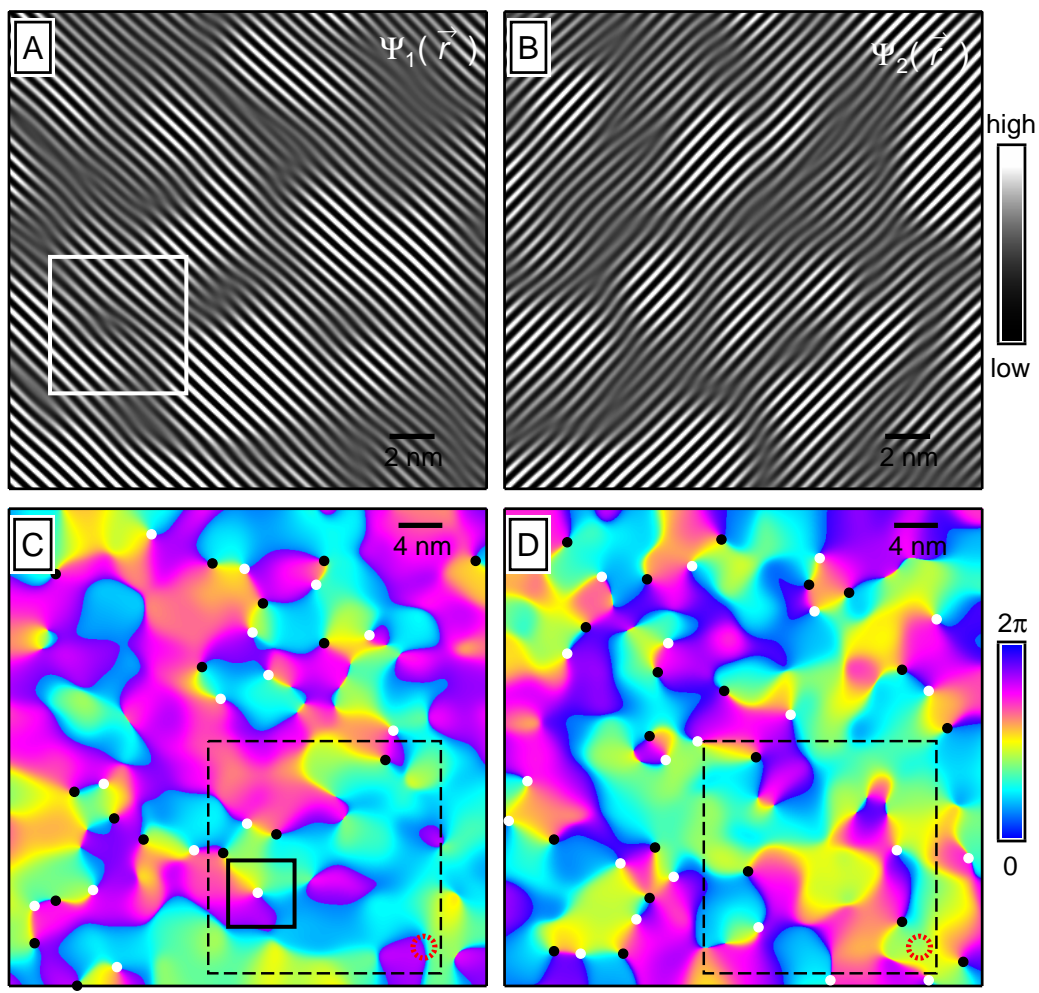

Figure 3

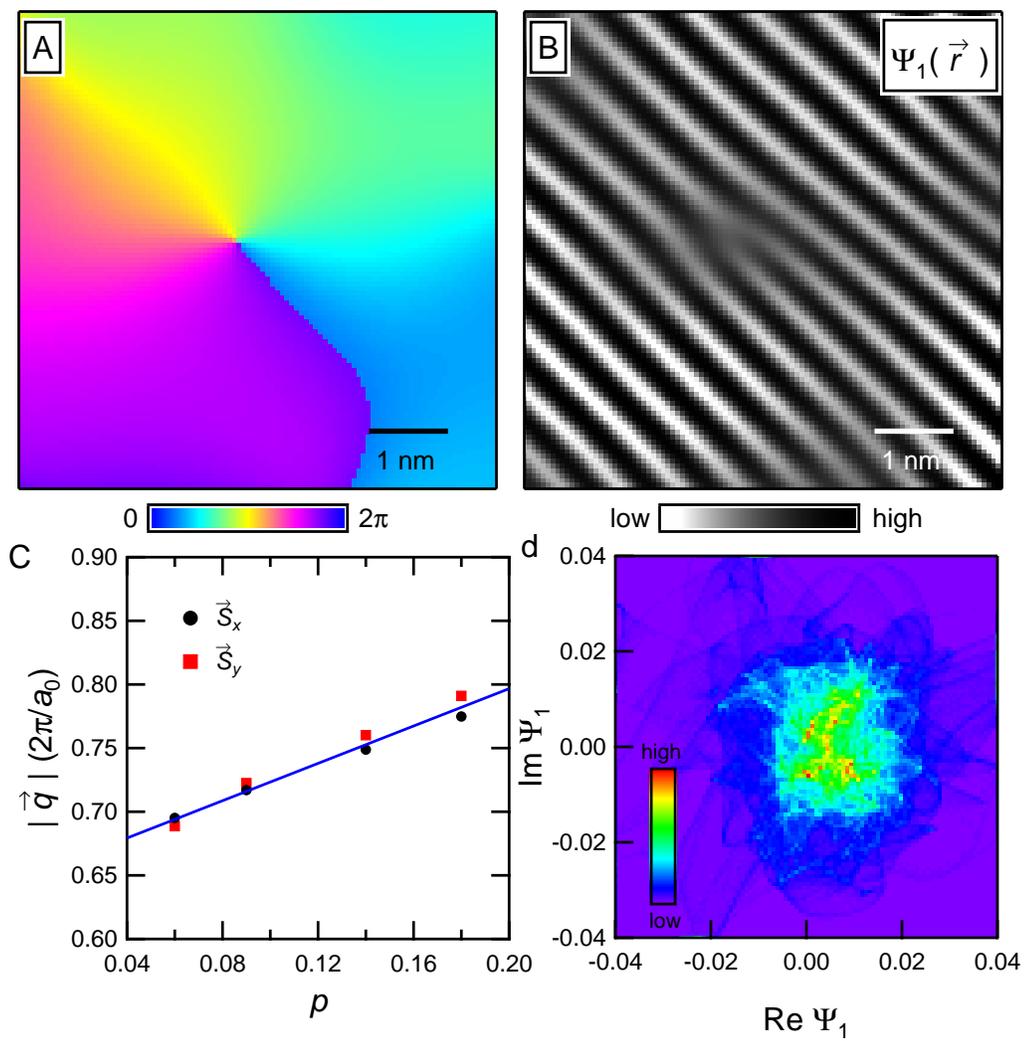

Figure 4

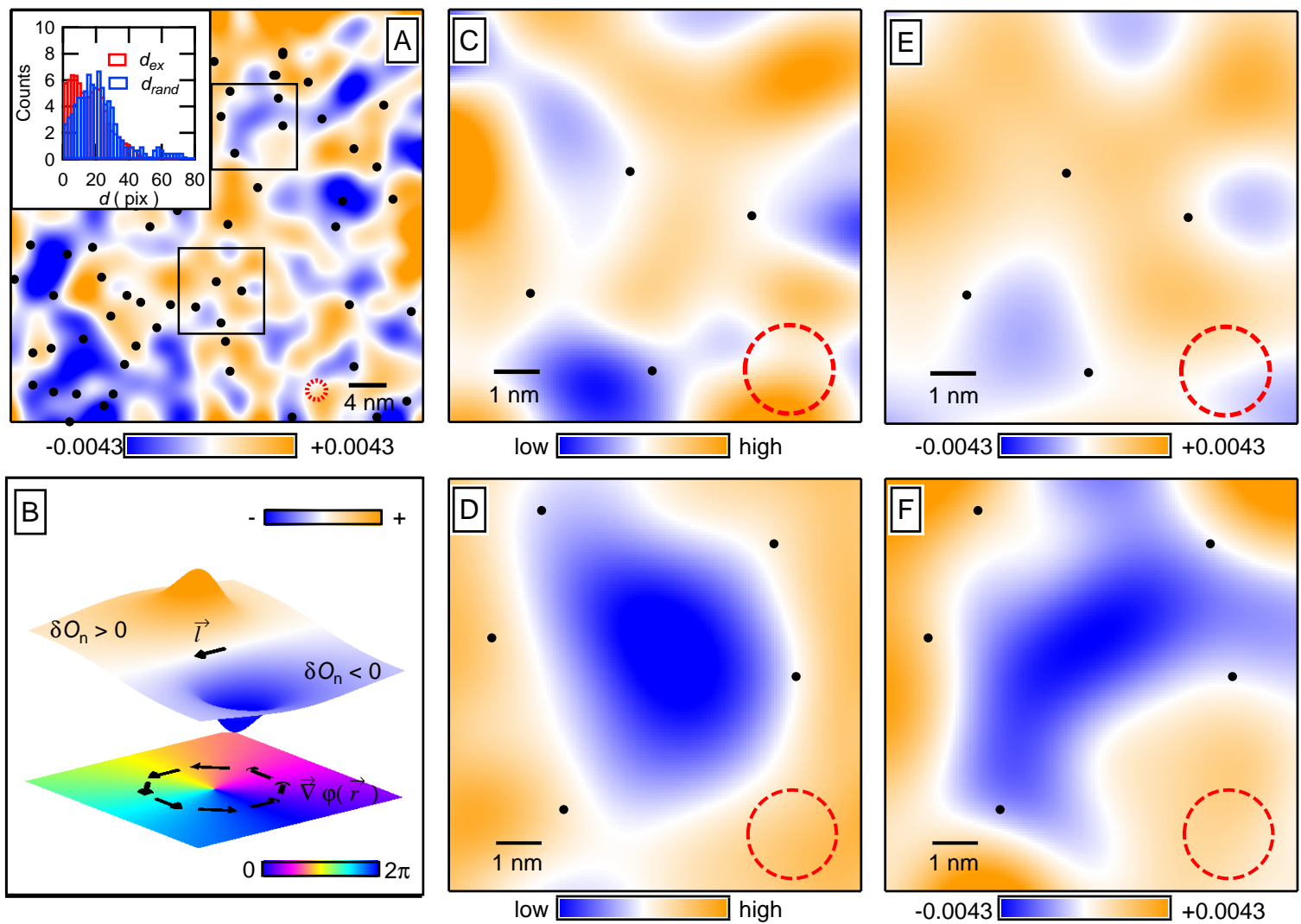

# Supporting Online Materials

# How Topological Defects Couple the Smectic Modulations and Intra-unit-cell Nematicity of the Cuprate Pseudogap States

A. Mesaros, K. Fujita, H. Eisaki, S. Uchida, J.C. Davis, S. Sachdev, J. Zaanen, M. J. Lawler, & Eun-Ah Kim*

*To whom correspondence should be addressed. E-mail: eun-ah.kim@cornell.edu

## Supporting Text and Figures

**(a) Procedure for measuring $Z(\vec{r}, E)$**

The STM tip-sample differential conductance $dI/dV(\vec{r}, E = eV)$ is measured at spatial locations $\vec{r}$ with sub-unit-cell resolution and at electron energies $E$. The "Z-map" technique is crucial because it cancels the otherwise profound and unavoidable systematic errors in $dI/dV(\vec{r}, E)$ alone. At low energy, in both the superconducting and pseudogap phases, only dispersive Bogoliubov quasiparticle-interference modulations are observed; these obviously cannot be cast in terms of smectic modulations (or dislocations) because the $dI/dV(\vec{r}, E)$ or $Z(\vec{r}, E)$ are quite different at each $E$.

**(b) Procedure for generating $O_n^Q(\vec{r}, e)$ and $O_s^Q(\vec{r}, e)$**

To visualize the spatial structures contributing to $O_n^Q(e)$ we define a coarse grained field with a coarsening length scale $1/\Lambda_N$ which acts as an effective 'ultra violet' cutoff at the $\Lambda_N$ length scale (see circles around the Bragg peaks in Fig. 1b inset). Then

$$\tilde{Z}(\vec{Q}; \vec{r})_\Lambda \equiv \sum_{r'} Z(\vec{r}') e^{i\vec{Q} \cdot (\vec{r}' - \vec{u}(\vec{r}'))} f_\Lambda(\vec{r}' - \vec{r})$$

$$\approx \frac{1}{\sqrt{N}} \sum_{\vec{k}} \tilde{Z}(\vec{Q} - \vec{k}) e^{i\vec{k} \cdot (\vec{r} - \vec{u}(\vec{r}))} e^{-k^2 / 2\Lambda^2}$$

(S1.1)

where $f_\Lambda(\vec{r}) \equiv (\Lambda^2/2\pi) e^{-\Lambda^2|\vec{r}|^2/2}$ is used to implement the cutoff at length scale $1/\Lambda$ for nematicity, we set this cutoff to the $3\sigma$ radius of the Bragg peaks $\Lambda_N$ (Fig. 1b inset). A local image of the nematicity $O_n^Q(\vec{r},e)$ is then given by:

$$O_n^Q(\vec{r},e) \propto \tilde{Z}(\vec{Q}_x;\vec{r})_{\Lambda_N} - \tilde{Z}(\vec{Q}_y;\vec{r})_{\Lambda_N} + \tilde{Z}(-\vec{Q}_x;\vec{r})_{\Lambda_N} - \tilde{Z}(-\vec{Q}_y;\vec{r})_{\Lambda_N} \Big|_e \quad (S1.2)$$

with the normalization requiring the field of view (FOV) average to equal $O_n^Q$, i.e., $\langle O_n^Q(\vec{r}) \rangle = O_n^Q$. Similarly, a local image of any smecticity $O_s^Q(\vec{r},e)$ is given by

$$O_s^Q(\vec{r},e) \propto \tilde{Z}(\vec{S}_x;\vec{r})_{\Lambda_S} - \tilde{Z}(\vec{S}_y;\vec{r})_{\Lambda_S} + \tilde{Z}(-\vec{S}_x;\vec{r})_{\Lambda_S} - \tilde{Z}(-\vec{S}_y;\vec{r})_{\Lambda_S} \Big|_e \quad (S1.3)$$

where for the smectic modulations we set this cutoff to the $3\sigma$ radius of the $\vec{S}_x$, $\vec{S}_y$ peaks $\Lambda_S$ (Fig. 1b inset), and again normalized to $\langle O_s^Q(\vec{r}) \rangle = O_s^Q$.

**(c) Statistical validation of the topological defects located near the $\delta O_n(\vec{r},e=1)=0$ contour**

A $\delta O_n(\vec{r},e=1)=0$ contour of Fig. 4a is shown in Fig. S1a together with topological defects(solid red circles). Each topological defect is linked to the nearest location on the $\delta O_n(\vec{r},e=1)=0$ contour marked by black solid circles. Distance $d_{ex}$ between each topological defects and their nearest location on the $\delta O_n(\vec{r},e=1)=0$ contour is calculated and a distribution of $d_{ex}$ is shown in Fig. S1c. In order to validate that the topological defects are located near the $\delta O_n(\vec{r},e=1)=0$ contour, same number of defects are randomly generated and distance $d_{rand}$ is calculated in a same way. Fig. S1b shows an example for the configuration of the randomly generated defects (solid blue circles). As shown in Fig. S1c, the distribution of $d_{rand}$ deviates significantly from $d_{ex}$ at short distances: while $d_{ex}$ distribution is peaked at near zero value $d_{rand}$ is peaked at a finite value. This contrast in the distance distribution indicates that the configuration of the topological defects found near the $\delta O_n(\vec{r}, e = 1) = 0$ contour is significantly different and is unlikely to be a part of configuration for the randomly generated defects.

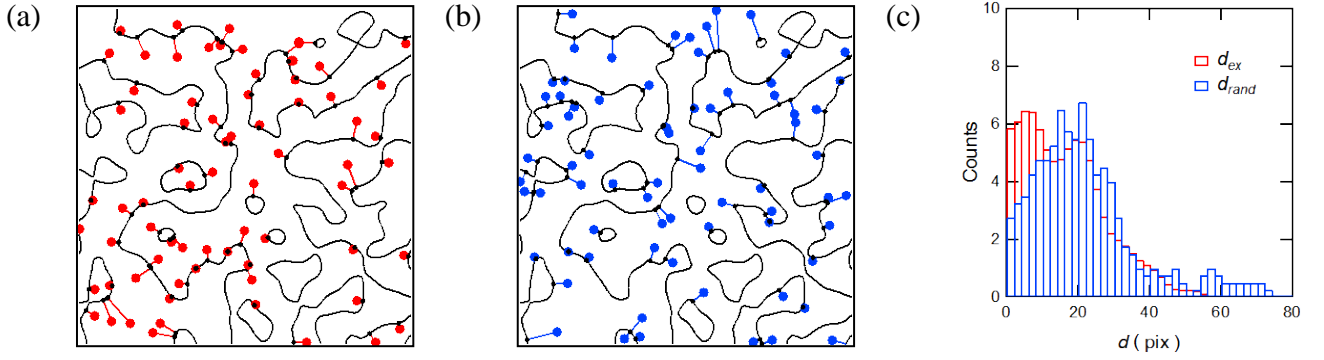

**Figure S1.** (a) $\delta O_n(\vec{r}, e = 1) = 0$ contour(solid lines) and topological defects(solid red). Each topological defect is linked to the nearest location on the $\delta O_n(\vec{r}, e = 1) = 0$ contour (solid black). (b) Same as (a) but randomly generated defects (solid blue) case. (c) Distribution of distances in (a) and (b) in units of pixels.

### (d) GL theory for coupling of smectic and nematic electronic structure components

In order to discuss coupling between smectic and nematic we first note the relevant symmetry and appropriate continuum fields. The long range nematicity present within the field of view [18] reduces the point group symmetry to an orthorhombic symmetry ($C_{2v}$). It also implies the fluctuation around the non-zero expectation value of the nematicity $\delta O_n(\vec{r}) \equiv O_n(\vec{r}) - \langle O_n \rangle$ is the appropriate field to represent nematicity in the GL functional. On the other hand, since long range smectic modulations is absent, smectic order parameter fields $\psi_1(\vec{r}), \psi_2(\vec{r})$ are themselves appropriate fields for expanding the GL functional. We start by considering the GL functional for the smectic modulations

$$F_s(\psi_1, \psi_2) = \int d^2\vec{r} \left[ a_{x,1} |\nabla_x \psi_1|^2 + a_{y,1} |\nabla_y \psi_1|^2 + m_1 |\psi_1|^2 + a_{x,2} |\nabla_x \psi_2|^2 + a_{y,2} |\nabla_y \psi_2|^2 + m_2 |\psi_2|^2 \right] \quad \text{(S2.1)}$$

This is the smectic functional introduced in Eq. (2) of main text, but written for both modulations ($s = 1,2$). Here the coupling constants $a_i$ and $m$ of Eq. (S2.1) are enhanced to $a_{i,s}$ and $m_s$ by an additional index $s$ labeling the corresponding smectic modulation. We kept terms that are allowed by the $C_{2v}$ symmetry, up to quadratic order in smectic fields.

The nematic fluctuation couples to the *phase* of smectic fields by locally shifting the smectic wavevectors:

$$\vec{S} \to \vec{S} + \vec{c}\,\delta O_n(\vec{r}). \tag{S2.2}$$

The vector $\vec{c}$ is a set of phenomenological coupling constants. For the coarse grained smectic field $\psi_s(\vec{r})$, this is equivalent to replacing the derivatives in Eq. (S2.1) by covariant derivatives:

$$\nabla_i \psi_s(\vec{r}) \to (\nabla_i + ic_i \delta O_n(\vec{r}))\psi_s(\vec{r}). \tag{S2.3}$$

As for the coupling to the *amplitude* of smectic fields, the lowest order term allowed by the symmetry is

$$\beta_s \delta O_n(\vec{r})|\psi_s(\vec{r})|^2, \tag{S2.4}$$

where $\beta_s$ for $s=1,2$ are two phenomenological constants. This term describes a local enhancement of smectic amplitude fluctuation caused by the nematic fluctuation.

We arrive at the final form of the GL functional taking the substitution given in Eq. (S2.3) and including the term Eq. (S2.4) to the smectic functional of Eq. (S2.1) and adding the nematic functional $F_n[\delta O_n]$:

$$\begin{aligned}
F_{GL}[\delta O_n, \psi_1, \psi_2] &= F_n[\delta O_n] \\
&+ \int d^2\vec{r} \sum_{s=1,2} \left[ \alpha_{x,s} |(\nabla_x + ic_x \delta O_n)\psi_s|^2 + \alpha_{y,s} |(\nabla_y + ic_y \delta O_n)\psi_s|^2 + m_s |\psi_s|^2 + \beta_s \delta O_n | \right. \\
&= F_n[\delta O_n] + F_s[\psi_1, \psi_2] \\
&+ \int d^2\vec{r} \sum_{s=1,2} \sum_{i=x,y} \left[ \alpha_{i,s} \delta O_n |\psi_s|^2 \nabla_i \varphi_s + \beta_s \delta O_n |\psi_s|^2 + \gamma_s \delta O_n^2 |\psi_s|^2 \right]
\end{aligned}$$
(S2.5)

The first line represents Equation (5) of main text, here written for both smectic modulations, with the added amplitude coupling from Eq. (S2.4) and $F_n[\delta O_n]$ is nematic functional to lowest order (quadratic) in $\delta O_n(\mathbf{r})$:

$$F_n[\delta O_n] = \int d^2\vec{r} \left[ \sum_{i=x,y} (\nabla_i \delta O_n)^2 + \frac{1}{\xi_N^2} \delta O_n^2 \right] \tag{S2.6}$$

where $\xi_N$ is the nematic fluctuation correlation length. In the second line of Eq. (S2.5), we have isolated the coupling terms and defined compact labels for the coupling constants $\alpha_{i,s} \equiv a_{i,s} c_i$ and $\gamma_s \equiv \sum_{i=x,y} a_{i,s} c_i^2$. $F_{GL}[\delta O_n, \psi_1, \psi_2]$ contains all terms up to quadratic order in each continuum field. Note that the existence of $\alpha_{x,2}$, $\alpha_{y,1}$ reflects absence of mirror symmetry with respect to yz plane or xz plane. Superlattice modulation breaks these mirror symmetries explicitly. Hence we did not assume these mirror symmetries in our GL functional. In the rest, we ignore $\gamma_s$ terms in comparison to $\beta_s$ terms.

**(e) The footprint of the GL functional in the vicinity of a single topological defect**

Here we discuss the key signature of the coupled GL functional of Eq. (5) depicted in Figure 4 based on the mean field theory treatment of the coupling. We are particularly interested in the coupling between a smectic topological defect and the nematicity. Assuming the spatial dependence of $\psi_s$ to be mostly due to the phase fluctuations away from the defect cores, i.e., $\psi_s \equiv |\psi_s| e^{i\varphi_s(\vec{r})}$ with $|\psi_s|$ constant, the saddle point equation for minimizing the GL functional of Eq. (S2.5) with respect to variations in $\delta O_n$ is

$$\left( -\nabla^2 + \xi_N^{-2} + 2\sum_{s=1,2} \vec{\alpha}_s \cdot \vec{c} |\psi_s|^2 \right) \delta O_n(\vec{r}) + \sum_{s=1,2} [2\vec{\alpha}_s \cdot \vec{\nabla}\varphi_s + \beta_s] |\psi_s|^2 = 0. \tag{S3.1}$$

Since $|\psi_s|$ is a constant, we can define a renormalized correlation length $\tilde{\xi}_N$ and rewrite Eq S3.1 as

$$\left(-\nabla^2 + \tilde{\xi}_N^{-2}\right)\delta O_n(\vec{r}) = -\sum_{s=1,2}[2\vec{\alpha}_s \cdot \vec{\nabla}\varphi_s + \beta_s]|\psi_s|^2.$$
(S3.2)

Much insight can now be gained by focusing on the vicinity of a single topological defect. If we consider a defect in $\psi_1(\vec{r})$, $\vec{\nabla}\varphi_2(\vec{r}) = 0$. Now Eq (S3.2) becomes

$$\left(-\nabla^2 + \tilde{\xi}_N^{-2}\right)\delta O_n(\vec{r}) = -2\vec{\alpha}_1 \cdot \vec{\nabla}\varphi_1(\vec{r})|\psi_1|^2 - \sum_{s=1,2}\beta_s|\psi_s|^2.$$
(S3.3)

To understand the solution of (S3.3), first consider the long distance limit $|\vec{r}| \gg \tilde{\xi}_N$ where the Laplacian can be neglected next to $\tilde{\xi}_N^{-2}$. In this limit, Eq (S3.3) can be solved by

$$\delta O_n = \vec{l}_1 \cdot \vec{\nabla}\varphi_1 - \tilde{\xi}_N^2 \sum_{s=1,2}\beta_s|\psi_s|^2,$$
(S3.4)

with

$$\vec{l}_1 \equiv -2\tilde{\xi}_N^2 |\psi_1|^2 \vec{\alpha}_1$$
(S3.5)

When the amplitude coupling $\beta_s|\psi_s|^2$ terms are left out, the solution Eq (S3.4) becomes the form quoted in the main text. In order to understand what this means for the distance between the defect location and the line of $\delta O_n(\vec{r}) = 0$, let us first consider the absence of the $\beta_s|\psi_s|^2$ terms and a simple uniform phase winding model of the defect with $\varphi_1(\vec{r}) = \theta(\vec{r})$ and $\vec{\nabla}\varphi_1(\vec{r}) = \frac{1}{|\vec{r}|}\hat{\theta}(\vec{r})$, where the polar coordinates $\vec{r} = (|\vec{r}|, \theta(\vec{r}))$ are centered on the defect and the polar angle $\theta(|\vec{r}|) = \tan^{-1}(r_x/r_y)$. For such a defect, $|\vec{r}| \cdot \vec{\nabla}\varphi(\vec{r}) = 0$ for all $\vec{r}$. There are two points on the circle of radius $r$ at which $\vec{r} // \pm \vec{l}$ for the given $\vec{l}_s$, and for these values of $\vec{r}$, $\vec{l}_s \cdot \vec{\nabla}\varphi(\vec{r}) = \vec{r} \cdot \vec{\nabla}\varphi(\vec{r}) = 0$ so that $\delta O_n(\vec{r}) = 0$. Moreover, this requires $\delta O_n(\vec{r})$ to change the sign on either side of the radial vector along the $\vec{l}_s$ direction since $\vec{l}_s \cdot \vec{\nabla}\varphi(\vec{r})$ change sign when the circle of radius $r$ crosses the line through the origin that is parallel to $\vec{l}_s$ (see Figure 4b). Now the effect of the $\beta_s|\psi_s|^2$ terms is to shift the value of $\delta O_n(\vec{r})$ uniformly. This will shift the line of vanishing $\delta O_n(\vec{r})$ away from the straight line through the origin that is parallel to $\vec{l}_s$, and bend the line into a curve. However, so long as the amplitude couplings are negligible compared to the phase coupling, i.e. $\left|2\vec{\alpha}_1 \cdot \vec{\nabla}\varphi_1(\vec{r})|\psi_1|^2\right| \gg \sum_{s=1,2}\beta_s|\psi_s|^2$, $\beta_s$ terms do not influence the robust feature of the close proximity between the defect location and the line of vanishing $\delta O_n(\vec{r})$. Moreover, since $\delta O_n(\vec{r})$ is defined as the deviation away from the non-zero average $O_n$, the overall shift due to the amplitude coupling can be absorbed into the definition of the average. Hence although the amplitude coupling is more relevant in the RG sense compared to the phase coupling through derivative, it is redundant.

The above reasoning can be verified in a more detailed calculation. The full solution to the saddle point equation (S3.4) can be obtained using the Green's function method

$$\delta O_n(\vec{r}) = -\int d^2\vec{r}'G(\vec{r}-\vec{r}')\left[2\vec{\alpha}_1 \cdot \vec{\nabla}\varphi_1(\vec{r}')|\psi_1(\vec{r}')|^2 + \sum_{s=1,2}\beta_s|\psi_s(\vec{r}')|^2\right]. \tag{S3.6}$$

where $G(\vec{r}) = K_0(|\vec{r}|/\tilde{\xi}_N)$, with $K_0(x)$ the modified-Bessel function of the second kind. Figure S1 shows the resulting $\delta O_n(\vec{r})$ configuration in the vicinity of a single topological defect. The defect is modeled as a phase winding with the amplitude suppression near the core,

$$\psi_1(\vec{r}) = \sqrt{(1-\exp(-|\vec{r}|^2/\xi_S^2))}e^{i\theta(\vec{r})}, \tag{S3.7}$$

where $\theta(\vec{r}) = \tan^{-1}(r_x/r_y)$, for the choice of $\vec{\alpha}_1 \equiv (\alpha_{x,1},\alpha_{y,1}) = (1,0)$ and two different values of $\beta$: $\beta=0$ (Figure S2a) and $\beta=1$ (Figure S2b). Here $\xi_S$ is the coherence length of the smectic. For such defect configuration, Eq(S3.6) reduces to Eq(S3.4) when $|\vec{r}| \gg \tilde{\xi}_N, \xi_S$.

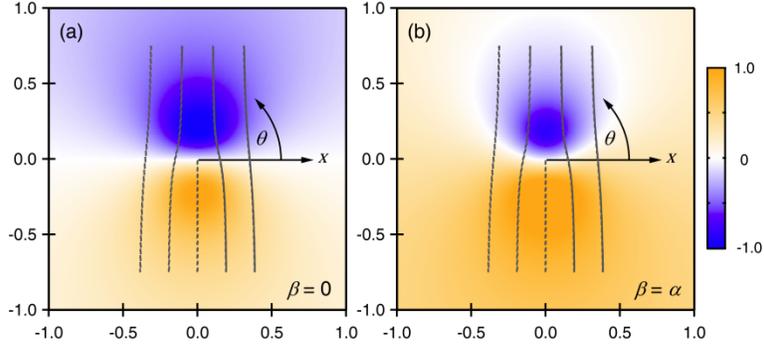

**Figure S2**: $\delta O_n(\vec{r})$ configuration calculated from Equation (S3.6) and (S3.7) for $\vec{a}_1 \equiv (a_{x,1}, a_{y,1}) = (1,0)$. (a) In the absence of the amplitude coupling setting $\beta_1 = 0$. Dashed lines schematically represent the peaks of the smectic order ("stripes"). This is the top view of Figure 4b. The fluctuation boundary is along the $\vec{a}_1 \equiv (a_{x,1}, a_{y,1}) = (1,0)$ direction. (b) The coupling to smectic amplitude $\beta_s$ only provides an overall shift of the pattern and introduces a curvature the fluctuation boundary passing through the defect.

In Figure S2a, the resulting $\delta O_n(\vec{r})$ vanishes along the $x$-axis which is the direction of $\vec{l}_1 \parallel \vec{\alpha}_1$, in agreement with the observation made above for the long distance limit. This corresponds to the top view of the main Figure 4b. This is consistent with the fact that the topological defects are mostly distributed on top of the curve defined by $\delta O_n(\vec{r}) = 0$ as indicated in the main Figure 4a. Figure S2b shows that the amplitude coupling (the $\beta_s$ term in Equation (5)) provides an overall shift of the source field, and therefore does not influence these robust features, unless $\beta_s \gg \alpha_{i,s}$.

## (f) Reproduction of nematic fluctuation based on topological defects within smectic fields

We use equation (S3.6) to simulate the nematic fluctuation $\delta O_n(\vec{r})$ using the experimentally obtained smectic fields $\psi_1(\vec{r})$ and $\psi_2(\vec{r})$. To do this we need to estimate the relative amplitudes of the GL phenomenological coupling constants $\alpha_{i,s}$ and $\beta_s$. They are searched in the large number of parameter space, and we found the best combination of them that provides the highest cross-correlation between $\delta O_n(\vec{r})$ and $\delta O_n^{sim}(\vec{r})$, as listed in the table below. That a phase coupling is larger than amplitude couplings when the amplitude coupling is more relevant in the RG sense shows that short distance physics is dominating and system is in a phase rather than a critical regime.

|  | $\alpha_{x,1}$ | $\alpha_{x,2}$ | $\alpha_{y,1}$ | $\alpha_{y,2}$ | $\beta_1$ | $\beta_2$ | $C$ |
|---|---|---|---|---|---|---|---|
| Fig. 4c | 4 | 16 | 4 | -4 | 8 | 2 | 0.56 |
| Fig. 4d | 0 | -12 | 4 | 0 | -10 | -4 | 0.62 |

**Table S1.** The list of nematic-smectic GL coupling terms of eq. (S3.6) and cross correlation coefficient. The "high statistical significance" cutoff for |C| is 20%.

Statistical significance of the cross correlation coefficient $C$ between two independent images can be assessed by calculating the probability $P(C, N_{eff})$,

$$P(C, N_{eff}) = B\left(\frac{1}{2}, \frac{N_{eff}-2}{2}\right)(C^2)^{-1/2}(1-C^2)^{N_{eff}/2-2}, \tag{S4.14}$$

where $B$ is the Beta function and $N_{eff}$ is the effective resolution defined by the number of pixel $N$ divided by $\pi\Gamma^2$. $\Gamma$ is a width of the Gaussian used for smoothing the fields, namely, $N_{eff} \equiv N/\pi\Gamma^2$. A criterion of significance can be then stated for $C$: If, for a given $N_{eff}$ and $C$ of our datasets, there is more than 5% probability that completely independent datasets exhibit the same or higher correlation than $C$, i.e. $P(C, N_{eff}) > 5\%$, then the observed correlation is not statistically significant. The $C$ in that case should be regarded as zero. In case of our datasets, this criterion says that cross-correlation having $|C| \geq 20\%$ are statistically significant. In fact, obtained cross-correlation coefficients between $\delta O_n(\vec{r})$ and $\delta O_n^{sim}(\vec{r})$ are greater than 50% as listed in the table S1.

**(g) Influence of electronic disorder on the topological defects**

In Fig. S3, smectic $2\pi$ topological defects are located in an image of simultaneously measured $\Delta_1(\vec{r})$. It is obvious that the topological defects can be found near the boundary between the relatively high and small $\Delta_1(\vec{r})$ regions. This reminds us the presence of the dopant oxygen sitting near the apical oxygen outside the $CuO_2$ plane since they are known to be identified in the relatively large $\Delta_1(\vec{r})$ regions.

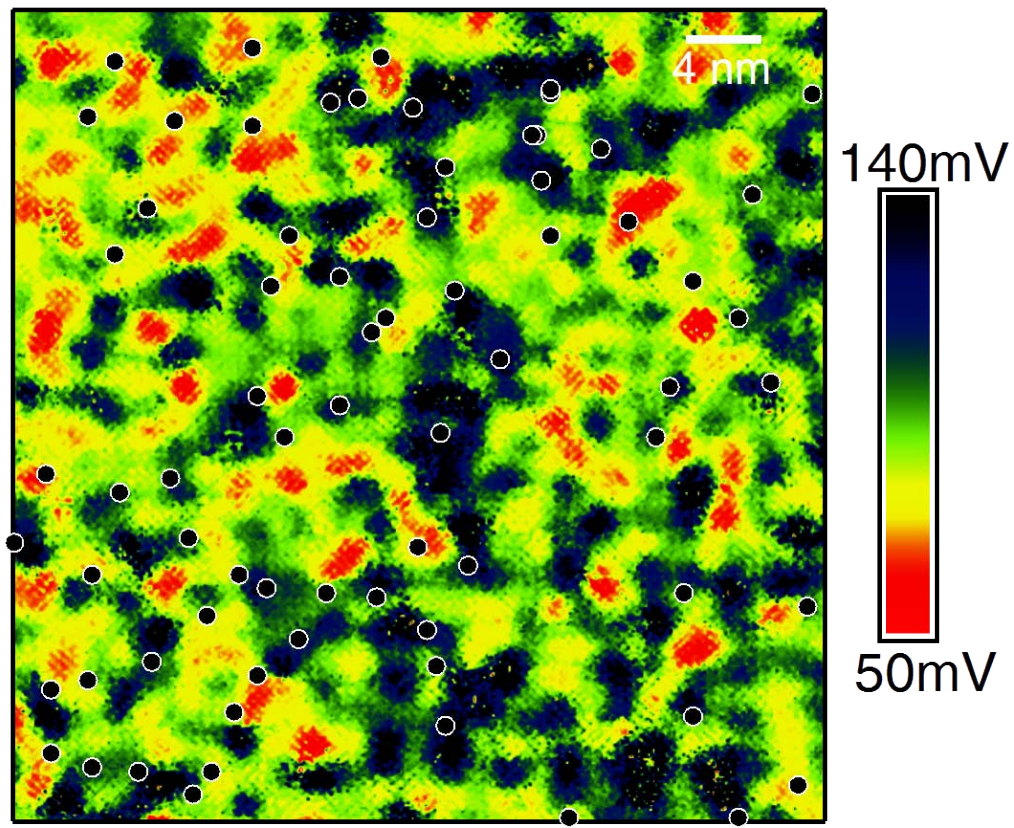

**Figure S3.** $\Delta_1(\vec{r})$ with simultaneously measured locations of smectic topological defects shown as black dots.